\documentclass[12pt,prd,showpacs,tightenlines,nofootinbib]{revtex4}
\usepackage{bm}
\usepackage{graphics}
\usepackage{rotating}
\usepackage{epsfig}
\begin{document}
\begin{flushright}{HU-EP-08/67}\end{flushright}
\title{Masses of light tetraquarks and scalar mesons in the relativistic quark model}
\author{D. Ebert$^{1}$, R. N. Faustov$^{2}$  and V. O. Galkin$^{1,2}$}
\affiliation{
$^1$ Institut f\"ur Physik, Humboldt--Universit\"at zu Berlin,
Newtonstr. 15, D-12489  Berlin, Germany\\
$^2$ Dorodnicyn Computing Centre, Russian Academy of Sciences,
  Vavilov Str. 40, 119991 Moscow, Russia}

\begin{abstract}
Masses of the ground state light tetraquarks are dynamically calculated in the
framework of the relativistic diquark-antidiquark picture. The
internal structure of the diquark is taken into account by calculating the
form factor of the diquark-gluon interaction in terms of the overlap
integral of the diquark wave functions. It is found that 
scalar mesons with  masses below 1 GeV:  $f_0(600)$ ($\sigma$), $K^*_0(800)$ ($\kappa$), 
$f_0(980)$ and $a_0(980)$ agree well with the light tetraquark interpretation.
\end{abstract}


\maketitle
The  consistent theoretical understanding of the light
meson sector remains an important problem already for many
years.~\footnote{A vast literature on the light meson
  spectroscopy is available. Therefore we mostly refer to the recent
  reviews where the references to earlier review and original papers
  can be found.} An extensive analysis of the data on highly excited 
light  non-strange meson states up to a mass of 2400 MeV collected by
Crystal Barrel experiment at LEAR (CERN)  has been published
\cite{bugg}. Classification of these new data requires better
theoretical description of light meson mass spectra. This is especially
important, since light exotic states (such as tetraquarks, glueballs,
hybrids)  within quantum chromodynamics (QCD) are expected to have masses
in this range \cite{at,kz,glb}.  
Particular interest is focused on scalar mesons, their
properties and abundance. A generally accepted consistent picture has
not yet emerged. Experimental and theoretical evidence \cite{kz} for the
existence of $f_0(600)(\sigma)$, $K^*_0(800)(\kappa)$, $f_0(980)$ and
$a_0(980)$ indicates that lightest scalars form a
full $SU(3)$ flavour nonet.  A peculiar feature of their mass
spectrum is the inversion of the  mass ordering, which cannot be
naturally understood in the 
$q\bar q$ picture. This fact stimulated various alternative
interpretations of light scalars as four quark states (tetraquarks)
\cite{jaffe,achasov} in particular
diquark-antidiquark bound states \cite{maiani}. 
The proximity of
$f_0/a_0$ to the $K\bar K$ threshold  led to the $K\bar K$ molecular
picture \cite{molec}. 

In this paper we calculate the masses of the ground state ($\langle
{\bf L}^2\rangle$=0) light tetraquarks as diquark-antidiquark bound
states in the relativistic quark model based on the quasipotential
approach in quantum chromodynamics. 
Following \cite{jaffe,maiani} the
diquark is taken in the colour antitriplet state.
Recently, in the framework of the same
model \cite{tetr1,tetr2} we investigated  the mass spectra of  
heavy tetraquarks. It was found that many of the newly observed
charmonium-like states \cite{pakhlova} above  open charm threshold, including
explicitly exotic charged states, could be
interpreted as tetraquark states with hidden charm.
In the present analysis of light tetraquarks we 
use,  as previously,
the  diquark-antidiquark 
picture to reduce a complicated relativistic 
four-body problem to the subsequent two more simple two-body
problems. The first step involves
the calculation of the masses, wave
functions and form factors of the diquarks, composed from light 
quarks. At the second step, a light tetraquark is considered to be a
bound diquark-antidiquark system. It is 
important to emphasize that we do not consider the diquark as a point
particle but explicitly take into account its structure
given by
the form factor of the diquark-gluon interaction in terms of the
diquark wave functions.

In the quasipotential approach and diquark-antidiquark picture of
tetraquarks the interaction of two quarks in a diquark and
the diquark-antidiquark interaction in a tetraquark are described
by the diquark wave function ($\Psi_{d}$) of the bound quark-quark
state and by the tetraquark wave function ($\Psi_{T}$) of the
bound diquark-antidiquark state, respectively. These wave functions satisfy the
quasipotential equation of the Schr\"odinger type \cite{efg}
\begin{equation}
\label{quas}
{\left(\frac{b^2(M)}{2\mu_{R}}-\frac{{\bf
p}^2}{2\mu_{R}}\right)\Psi_{d,T}({\bf p})} =\int\frac{d^3 q}{(2\pi)^3}
 V({\bf p,q};M)\Psi_{d,T}({\bf q}),
\end{equation}
where the relativistic reduced mass is
\begin{equation}
\mu_{R}=\frac{E_1E_2}{E_1+E_2}=\frac{M^4-(m^2_1-m^2_2)^2}{4M^3},
\end{equation}
and $E_1$, $E_2$ are given by
\begin{equation}
\label{ee}
E_1=\frac{M^2-m_2^2+m_1^2}{2M}, \quad E_2=\frac{M^2-m_1^2+m_2^2}{2M}.
\end{equation}
Here, $M=E_1+E_2$ is the bound-state mass (diquark or tetraquark),
$m_{1,2}$ are the masses of quarks ($q=u,d$ and $s$) which form
the diquark or of the diquark ($d$) and antidiquark ($\bar d'$) which
form the light tetraquark ($T$), and ${\bf p}$ is their relative
momentum. In the center-of-mass system the relative momentum
squared on mass shell reads
\begin{equation}
{b^2(M) }
=\frac{[M^2-(m_1+m_2)^2][M^2-(m_1-m_2)^2]}{4M^2}.
\end{equation}

The kernel $V({\bf p,q};M)$ in Eq.~(\ref{quas}) is the
quasipotential operator of the quark-quark or diquark-antidiquark
interaction. It is constructed with the help of the off-mass-shell
scattering amplitude, projected onto the positive-energy states.
For the quark-quark interaction in a diquark we
use the relation $V_{qq}=V_{q\bar q}/2$ arising under the
assumption of an octet structure of the interaction from the
difference in the $qq$ and $q\bar q$ colour
states. An important role in this construction is played by the
Lorentz structure of the confining interaction. In our analysis of
mesons, while constructing the quasipotential of the
quark-antiquark interaction, we assumed that the effective
interaction is the sum of the usual one-gluon exchange term and a
mixture of long-range vector and scalar linear confining
potentials, where the vector confining potential contains the
Pauli term. We use the same conventions for the construction of
the quark-quark and diquark-antidiquark interactions in the
tetraquark. The quasipotential is then defined as follows
\cite{tetr1}.

(a) For the quark-quark  ($qq$, $qs$, $ss$) interactions,
$V({\bf p,q};M)$ reads
 \begin{equation}
\label{qpot}
V({\bf p,q};M)=\bar{u}_{1}(p)\bar{u}_{2}(-p){\cal V}({\bf p}, {\bf
q};M)u_{1}(q)u_{2}(-q),
\end{equation}
with
\[
{\cal V}({\bf p,q};M)=\frac12\left[\frac43\alpha_sD_{ \mu\nu}({\bf
k})\gamma_1^{\mu}\gamma_2^{\nu}+ V^V_{\rm conf}({\bf k})
\Gamma_1^{\mu}({\bf k})\Gamma_{2;\mu}(-{\bf k})+
 V^S_{\rm conf}({\bf k})\right].
\]
Here, $\alpha_s$ is the QCD coupling constant; $D_{\mu\nu}$ is the
gluon propagator in the Coulomb gauge,
\begin{equation}
D^{00}({\bf k})=-\frac{4\pi}{{\bf k}^2}, \quad D^{ij}({\bf k})=
-\frac{4\pi}{k^2}\left(\delta^{ij}-\frac{k^ik^j}{{\bf k}^2}\right),
\quad D^{0i}=D^{i0}=0,
\end{equation}
and ${\bf k=p-q}$; $\gamma_{\mu}$ and $u(p)$ are the Dirac
matrices and spinors,
\begin{equation}
\label{spinor}
u^\lambda({p})=\sqrt{\frac{\epsilon(p)+m}{2\epsilon(p)}}
\left(\begin{array}{c} 1\\
\displaystyle\frac{\mathstrut\bm{\sigma}\cdot{\bf p}}
{\mathstrut\epsilon(p)+m}
\end{array}\right)
\chi^\lambda,
\end{equation}
with $\epsilon(p)=\sqrt{{\bf p}^2+m^2}$.

The effective long-range vector vertex of the quark is
defined \cite{egf} by
\begin{equation}
\Gamma_{\mu}({\bf k})=\gamma_{\mu}+
\frac{i\kappa}{2m}\sigma_{\mu\nu}\tilde k^{\nu}, \qquad \tilde
k=(0,{\bf k}),
\end{equation}
where $\kappa$ is the Pauli interaction constant characterizing
the anomalous chromomagnetic moment of quarks. In configuration
space the vector and scalar confining potentials in the
nonrelativistic limit \cite{efg} reduce to
\begin{eqnarray}
V^V_{\rm conf}(r)&=&(1-\varepsilon)V_{\rm
conf}(r),\nonumber\\[1ex] V^S_{\rm conf}(r)& =&\varepsilon V_{\rm
conf}(r),
\end{eqnarray}
with
\begin{equation}
V_{\rm conf}(r)=V^S_{\rm conf}(r)+
V^V_{\rm conf}(r)=Ar+B,
\end{equation}
where $\varepsilon$ is the mixing coefficient.

(b) For the diquark-antidiquark ($d\bar d'$) interaction, 
 $V({\bf p,q};M)$ is given by

\begin{eqnarray}
\label{dpot} V({\bf p,q};M)&=&\frac{\langle
d(P)|J_{\mu}|d(Q)\rangle} {2\sqrt{E_dE_d}} \frac43\alpha_sD^{
\mu\nu}({\bf k})\frac{\langle \bar d'(P')|J_{\nu}|\bar d'(Q')\rangle}
{2\sqrt{E_{d'}E_{d'}}}\nonumber\\[1ex]
&&+\psi^*_d(P)\psi^*_{\bar d'}(P')\left[J_{d;\mu}J_{\bar d'}^{\mu} V_{\rm
conf}^V({\bf k})+V^S_{\rm conf}({\bf
k})\right]\psi_d(Q)\psi_{\bar d'}(Q'),
\end{eqnarray}
where $\langle
d(P)|J_{\mu}|d(Q)\rangle$ is the vertex of the
diquark-gluon interaction which takes into account the finite size of
the diquark 
$\Big[$$P^{(')}=(E_{d^{(')}},\pm{\bf p})$ and
$Q^{(')}=(E_{d^{(')}},\pm{\bf q})$,
$E_d=(M^2-M_{d'}^2+M_d^2)/(2M)$ and $E_{d'}=(M^2-M_d^2+M_{d'}^2)/(2M)$
$\Big]$.

The diquark state in the confining part of the diquark-antidiquark
quasipotential (\ref{dpot}) is described by the wave functions
\begin{equation}
 \label{eq:ps}
 \psi_d(p)=\left\{\begin{array}{ll}1 &\qquad \text{for a scalar
 diquark,}\\[1ex]
\varepsilon_d(p) &\qquad \text{for an axial-vector diquark,}
\end{array}\right.
\end{equation}
where the four-vector
\begin{equation}\label{pv}
\varepsilon_d(p)=\left(\frac{(\bm{\varepsilon}_d\cdot{\bf
p})}{M_d},\bm{\varepsilon}_d+ \frac{(\bm{\varepsilon}_d\cdot{\bf
p}){\bf
 p}}{M_d(E_d(p)+M_d)}\right), \qquad \varepsilon^\mu_d(p) p_\mu=0,
\end{equation}
is the polarization vector of the axial-vector diquark with
momentum ${\bf p}$, $E_d(p)=\sqrt{{\bf p}^2+M_d^2}$, and
$\varepsilon_d(0)=(0,\bm{\varepsilon}_d)$ is the polarization
vector in the diquark rest frame. The effective long-range vector
vertex of the diquark can be presented in the form
\begin{equation}
 \label{eq:jc}
 J_{d;\mu}=\left\{\begin{array}{ll}
 \frac{\displaystyle (P+Q)_\mu}{\displaystyle
 2\sqrt{E_dE_d}}&\qquad \text{ for a scalar diquark,}\\[3ex]
-\; \frac{\displaystyle (P+Q)_\mu}{\displaystyle2\sqrt{E_dE_d}}
 +\frac{\displaystyle i\mu_d}{\displaystyle 2M_d}\Sigma_\mu^\nu
\tilde k_\nu
 &\qquad \text{ for an axial-vector diquark.}\end{array}\right.
\end{equation}
Here, the antisymmetric tensor
$\Sigma_\mu^\nu$ is defined by
\begin{equation}
 \label{eq:Sig}
 \left(\Sigma_{\rho\sigma}\right)_\mu^\nu=-i(g_{\mu\rho}\delta^\nu_\sigma
 -g_{\mu\sigma}\delta^\nu_\rho),
\end{equation}
and the axial-vector diquark spin ${\bf S}_d$ is given by
$(S_{d;k})_{il}=-i\varepsilon_{kil}$; $\mu_d$ is the total
chromomagnetic moment of the axial-vector diquark.

The constituent quark masses 
$m_u=m_d=0.33$ GeV, $m_s=0.5$ GeV and the parameters of the linear
potential $A=0.18$ GeV$^2$ and $B=-0.3$~GeV, fixed previously \cite{egf}, have values
typical in quark models. The value of the mixing coefficient of
vector and scalar confining potentials $\varepsilon=-1$ has been
determined from the consideration of charmonium radiative decays
\cite{efg} and the heavy-quark expansion. The universal
Pauli interaction constant $\kappa=-1$ has been fixed from the
analysis of the fine splitting of heavy quarkonia ${ }^3P_J$ -
states \cite{efg}. In this case, the long-range chromomagnetic
interaction of quarks vanishes in accordance with the flux-tube
model.

At the first step, we take
the masses and form factors of the light
diquarks  from the previous consideration
of light diquarks in heavy baryons \cite{hbar}. 
 The form factor $F(r)$ entering the vertex of the
diquark-gluon interaction was expressed
through the overlap integral of the diquark wave functions. Our
estimates showed that this form factor can be approximated  with 
high accuracy by the expression 
\begin{equation}
  \label{eq:fr}
  F(r)=1-e^{-\xi r -\zeta r^2}.
\end{equation}
The values of the masses and parameters $\xi$ and $\zeta$ for light
scalar diquark $[\cdots]$ and axial vector diquark $\{\cdots\}$ ground states are
given in Table~\ref{tab:dmass},~\ref{tab:fcc}.

\begin{table}
  \caption{Masses of light ground state diquarks (in MeV). S and A
    denotes scalar and axial vector diquarks antisymmetric $[\dots]$ and
    symmetric $\{\dots\}$ in flavour, respectively. }
  \label{tab:dmass}
\begin{ruledtabular}
\begin{tabular}{ccccccc}
Quark& Diquark&  
\multicolumn{5}{c}{\hspace{-2.9cm}\underline{\hspace{5.1cm}Mass\hspace{5.1cm}}}
\hspace{-2.9cm} \\
content &type & \cite{hbar}& \cite{efkr}&\cite{burden}&\cite{maris} &
\cite{hess}\\
& &our &NJL &BSE & BSE &Lattice\\
\hline
$[u,d]$& S & 710 & 705 &737 &820& 694(22)\\
$\{u,d\}$& A & 909 & 875 &949 &1020&806(50)\\
$[u,s]$ & S& 948 & 895 &882&1100&\\
$\{u,s\}$& A & 1069 & 1050&1050&1300& \\
$\{s,s\}$& A & 1203 & 1215&1130&1440& \\
\end{tabular}
\end{ruledtabular}
\end{table}

\begin{table}
\caption{\label{tab:fcc}Parameters  $\xi$ and $\zeta$ for ground state
  light diquarks.}
\begin{ruledtabular}
\begin{tabular}{cccc}
Quark &Diquark& $\xi$  & $\zeta$  \\
content& type&(GeV)&(GeV$^2$)\\
\hline
$[u,d]$&S & 1.09 & 0.185  \\
$\{u,d\}$&A &1.185 & 0.365  \\
$[u,s]$& S & 1.23 & 0.225 \\
$\{u,s\}$& A & 1.15 & 0.325\\
$\{s,s\}$ & A& 1.13 & 0.280
\end{tabular}
\end{ruledtabular}
\end{table}

At the second step, we calculate the masses of light tetraquarks 
considered as the bound states of a light diquark and
antidiquark. For the
potential of the diquark-antidiquark interaction 
(\ref{dpot}) 
we get  in configuration space 
\cite{tetr2}  
\begin{eqnarray}
 \label{eq:pot}
 V(r)&=& \hat V_{\rm Coul}(r)+V_{\rm conf}(r)+\frac12\Biggl\{\left[
   \frac1{E_1(E_1+M_1)}+\frac1{E_2(E_2+M_2)}\right]
\frac{\hat V'_{\rm Coul}(r)}r 
-\Biggl[\frac1{M_1(E_1+M_1)}\nonumber\\[1ex]
&& +\frac1{M_2(E_2+M_2)}\Biggr]
\frac{V'_{\rm conf}(r)}r +\frac{\mu_d}2
\left(\frac1{M_1^2}+\frac1{M_2^2}\right)\frac{V'^V_{\rm conf}(r)}r\Biggr\}
{\bf L}\cdot ({\bf
S}_1+{\bf S}_2 )\nonumber\\[1ex]
&&+\frac12\Biggl\{\left[
   \frac1{E_1(E_1+M_1)}-\frac1{E_2(E_2+M_2)}\right]
\frac{\hat V'_{\rm Coul}(r)}r 
-\left[\frac1{M_1(E_1+M_1)}-\frac1{M_2(E_2+M_2)}\right]\nonumber\\[1ex]
&& \times
\frac{V'_{\rm conf}(r)}r +\frac{\mu_d}2
\left(\frac1{M_1^2}-\frac1{M_2^2}\right)\frac{V'^V_{\rm conf}(r)}r\Biggl\}
{\bf L}\cdot ({\bf
S}_1-{\bf S}_2 )\nonumber\\[1ex]
&&+\frac1{E_1E_2}\Biggl\{{\bf
 p}\left[\hat V_{\rm Coul}(r)+V^V_{\rm conf}(r)\right]{\bf p} -\frac14
\Delta V^V_{\rm conf}(r)+ \hat V'_{\rm Coul}(r)\frac{{\bf
 L}^2}{2r}\nonumber\\[1ex]
&& +\frac1{r}\left[\hat V'_{\rm
Coul}(r)+\frac{\mu_d}4\left(\frac{E_1}{M_1}
+\frac{E_2}{M_2}\right)V'^V_{\rm conf}(r)\right]{\bf L}\cdot ({\bf
S}_1+{\bf S}_2)\nonumber\\[1ex] &&
+\frac{\mu_d}4\left(\frac{E_1}{M_1}
-\frac{E_2}{M_2}\right)\frac{V'^V_{\rm conf}(r)}{r}{\bf
L}\cdot({\bf S}_1-{\bf S}_2)\nonumber\\[1ex] &&
+\frac13\left[\frac1{r}{\hat V'_{\rm Coul}(r)}-\hat V''_{\rm
Coul}(r) +\frac{\mu_d^2}4\frac{E_1E_2}{M_1M_2}
\left(\frac1{r}{V'^V_{\rm conf}(r)}-V''^V_{\rm
 conf}(r)\right)\right]\nonumber\\[1ex]
&&\times
 \left[\frac3{r^2}({\bf S}_1\cdot{\bf r}) ({\bf
 S}_2\cdot{\bf r})-
{\bf S}_1\cdot{\bf S}_2\right]\nonumber\\[1ex] &&
+\frac23\left[\Delta \hat V_{\rm
Coul}(r)+\frac{\mu_d^2}4\frac{E_1E_2}{M_1M_2} \Delta V^V_{\rm
conf}(r)\right]{\bf S}_1\cdot{\bf S}_2\Biggr\},
\end{eqnarray}
where $$\hat V_{\rm Coul}(r)=-\frac{4}{3}\alpha_s
\frac{F_1(r)F_2(r)}{r}$$ is the Coulomb-like one-gluon exchange
potential which takes into account the finite sizes of the diquark
and antidiquark through corresponding form factors $F_{1,2}(r)$.
Here, ${\bf S}_{1,2}$ and ${\bf L}$ are the spin operators of
diquark and antidiquark and the operator of the relative orbital
angular momentum.  In the
following we choose the total chromomagnetic moment of the
axial-vector diquark $\mu_d=0$. Such a choice appears to be
natural, since the long-range chromomagnetic interaction of
diquarks proportional to $\mu_d$ then also vanishes in accordance
with the flux-tube model.

We substitute the diquark-antidiquark interaction potential
(\ref{eq:pot}) in the wave equation (\ref{quas}) and solve
it numerically in configuration space. The calculated masses of the
ground states of light 
tetraquarks considered as  light diquark-antidiquark bound systems are
given in Tables~\ref{tab:lmass} and \ref{tab:kmass}. In
Table~\ref{tab:lmass} we present masses for light unflavoured
tetraquarks (without or with hidden strangeness) and in
Table~\ref{tab:kmass} - masses of strange tetraquarks. Possible
experimental candidates for such states are also given.

In the diquark-antidiquark picture of tetraquarks
both scalar $S$ (antisymmetric in flavour
$[\dots]$) and axial vector $A$ (symmetric in flavour
$\{\dots\}$) diquarks are considered. Therefore we get the
following structure of the light tetraquark  ground ($1S$) states
($C$ is defined only for neutral self-conjugated states): 
\begin{itemize}
\item Two states with $J^{PC}=0^{++}$:
\begin{eqnarray*}
&&X(0^{++})=S\bar S\\
&&X(0^{++}{}')=A\bar A
\end{eqnarray*}
\item Three states with 
$J^{PC}=1^{+\pm}$:
\begin{eqnarray*}
&&X(1^{++})=\frac1{\sqrt{2}}(S\bar A +\bar S A)\\
&&X(1^{+-})=\frac1{\sqrt{2}}(S\bar A-\bar S A)\\
&&X(1^{+-}{}')=A\bar A
\end{eqnarray*}
\item  One state with $J^{PC}=2^{++}$:
$$X(2^{++})=A\bar A.$$
\end{itemize}
The lightest $S\bar S$ scalar ($0^{++}$) tetraquark states form the
SU(3) flavour nonet: one tetraquark 
($[ud][\bar u\bar d]$) with neither open or hidden
strangeness (electric charge $Q=0$ and isospin $I=0$); four
tetraquarks ($[sq][\bar u\bar d]$, $[\bar s\bar q][ud]$, $q=u,d$)  with open
strangeness ($Q=0,\pm 1$, $I=\frac12$) and four tetraquarks ($[sq][\bar
s\bar q']$) with hidden strangeness ($Q=0,\pm 1$, $I=0,1$).  Since we
neglect  in our model the mass difference of $u$ and 
$d$ quarks and electromagnetic interactions, the corresponding tetraquarks
will be degenerate in mass. 

\begin{table}
  \caption{Masses of light unflavored diquark-antidiquark ground state
    ($\langle {\bf L}^2\rangle$=0)
    tetraquarks (in
    MeV) and possible experimental candidates. S and A
    denote scalar and axial vector diquarks. }
  \label{tab:lmass}
\begin{ruledtabular}
\begin{tabular}{ccccccc}
State& Diquark &Theory&
\multicolumn{4}{l}{\underline{\hspace{3.4cm}Experiment \cite{pdg}\hspace{3.4cm}}} 
\hspace{-5.5cm} \\
$J^{PC}$ & content& mass & $I=0$&mass&$I=1$&mass  \\
\hline
$(qq)(\bar q\bar q)$\\
$0^{++}$ & $S\bar S$  & 596 & $f_0(600)$ ($\sigma$)& 400-1200&  &- \\
$1^{+\pm}$ & $(S\bar A\pm \bar S A)/\sqrt2$& 672& \\
$0^{++}$& $A\bar A$ & 1179 & $f_0(1370)$& 1200-1500& &  \\
$1^{+-}$& $A\bar A$ & 1773 & \\
$2^{++}$& $A\bar A$ & 1915 &
$\left\{\begin{array}{l}f_2(1910)\\f_2(1950) \end{array}\right.$&
$\left. \begin{array}{l}1903(9)\\1944(12)\end{array}\right.$& & \\
$(qs)(\bar q\bar s)$\\
$0^{++}$ & $S\bar S$  & 992 & $f_0(980)$ & 980(10) &$a_0(980)$  & 984.7(12)\\
$1^{++}$ & $(S\bar A +\bar S A)/\sqrt2$&
1201& $f_1(1285)$& 1281.8(6)&$a_1(1260)$ &1230(40)\\
$1^{+-}$ & $(S\bar A- \bar S A)/\sqrt2$&
1201& $h_1(1170)$& 1170(20)&$b_1(1235)$ &1229.5(32)\\
$0^{++}$& $A\bar A$ & 1480  & $f_0(1500)$& 1505(6)&$a_0(1450)$ &1474(19)\\
$1^{+-}$& $A\bar A$ & 1942 &$h_1(1965)$&1965(45)&$b_1(1960)$ &1960(35)\\
$2^{++}$& $A\bar A$ & 2097 & $\left\{\begin{array}{l}f_2(2010)\\f_2(2140)\end{array}\right.$ & $\left.\begin{array}{l}2011(70)\\2141(12)\end{array}\right.$&$\left\{\begin{array}{l}a_2(1990)\\a_2(2080)\end{array}\right.$ &$\left.\begin{array}{l}2050(45)\\2100(20)\end{array}\right.$ \\
$(ss)(\bar s\bar s)$\\
$0^{++}$& $A\bar A$ & 2203 & $f_0(2200)$& 2189(13)& &- \\
$1^{+-}$& $A\bar A$ & 2267 &$h_1(2215)$&2215(40)&&- \\
$2^{++}$& $A\bar A$ & 2357 & $f_2(2340)$ & 2339(60)& &- \\
 \end{tabular}
\end{ruledtabular}
\end{table}

\begin{table}
  \caption{Masses of strange diquark-antidiquark ground state
    ($\langle {\bf L}^2\rangle$=0) tetraquarks (in
    MeV) and possible experimental candidates. S and A
    denote scalar and axial vector diquarks. }
  \label{tab:kmass}
\begin{ruledtabular}
\begin{tabular}{ccccc}
State& Diquark &Theory&
\multicolumn{2}{l}{\underline{\hspace{1.2cm}Experiment \cite{pdg}\hspace{1.2cm}}} 
\hspace{-5.5cm} \\
$J^{P}$ & content&mass &$I=\frac12$& mass  \\
\hline
$(qq)(\bar s\bar q)$ or $(sq)(\bar q\bar q)$\\
$0^{+}$ & $S\bar S$  & 730 & $K^*_0(800)$ ($\kappa$)& 672(40)\\
$1^{+}$ & $(S\bar A\pm \bar S A)/\sqrt2$& 1057& \\
$0^{+}$& $A\bar A$ & 1332 & $K_0^*(1430)$& 1425(50)\\
$1^{+}$& $A\bar A$ & 1855 & \\
$2^{+}$& $A\bar A$ & 2001 & $K^*_2(1980)$ & 1973(26)\\
 \end{tabular}
\end{ruledtabular}
\end{table}

From  Tables~\ref{tab:lmass} and \ref{tab:kmass} we see that the
diquark-antidiquark picture can 
provide a natural explanation for the inversion of masses of light scalar $0^+$
mesons. Indeed all lightest experimentally
observed scalar mesons $f_0(600)$ ($\sigma$), $K^*_0(800)$ ($\kappa$), 
$f_0(980)$ and $a_0(980)$ can be interpreted in our model as light
tetraquarks composed from 
a
scalar diquark and
antidiquark ($S\bar S$). Therefore, the $f_0(980)$ and $a_0(980)$
tetraquarks contain, in comparison to 
the
$q\bar q$ picture, an additional pair of strange quarks which
gives a natural explanation why their masses are heavier than the strange
$K^*_0(800)$ ($\kappa$). Note that physical neutral scalar
states with $I=0$ are in fact mixtures of pure tetraquark $f_0$ and $\sigma_0$ states
\begin{eqnarray}
  \label{eq:mix}
  |f\rangle&=&\cos \varphi|f_0\rangle+\sin\varphi|\sigma_0\rangle,\cr
|\sigma\rangle&=&-\sin \varphi|f_0\rangle+\cos\varphi|\sigma_0\rangle,
\end{eqnarray}
where
$$f_0=\frac{1}{\sqrt2}([su][\bar
s \bar u]+[sd][\bar s\bar d]); \qquad\sigma_0=[ud][\bar u\bar
d].$$
However the $f-\sigma$ mixing is small because the Zweig rule is expected
to hold in the physical mass spectrum \cite{maiani}. Our results support such
conclusion,
 since we get good agreement with experimental data already
without such a mixing. 

The other scalar tetraquark states can be
composed from an axial vector diquark and antidiquark ($A\bar A$). Their
masses are predicted to be approximately 600 MeV heavier than the
$S\bar S$ tetraquarks.  The diquark-antidiquark composition also
naturally explains the experimentally observed proximity of masses of the
unflavored $a_0(1450)$, $f_0(1500)$ and strange $K_0^*(1430)$ scalars.  
Let us note that
quark-antiquark scalar states are predicted in
our model to
have masses around 1200 MeV ($q\bar q$) and 1400 MeV ($q\bar s$).  

The axial vector $1^+$ states can be composed both from scalar and
axial vector diquark and antidiquark ($(S \bar A\pm \bar S
A)/\sqrt{2}$) and from 
an axial vector diquark and antidiquark ($A\bar A$),
respectively.
 Our model predicts rather low mass values of the former states
composed from light quarks $(\{ud\}[\bar u\bar d]\pm\{\bar u\bar
  d\}[ud])/\sqrt2$, 672 MeV,
 and of their strange partner ($[qs]\{\bar u\bar
d\}\pm [\bar q\bar s]\{ud\}$),  1057 MeV. Such axial vector states are
not observed experimentally. Note that the recent study \cite{sg} also
indicates that corresponding light tetraquarks should have masses below 1
GeV. On the other hand, there are several
candidates for the axial vector $(\{qs\}[\bar q\bar s]\pm\{\bar q\bar
  s\}[us])/\sqrt2$ tetraquarks both in isospin $I=1$ ($a_1(1260)$,
$b_1(1235)$) and $I=0$ ($f_1(1285)$, $h_1(1170)$) channels. However,
ordinary $q\bar q$ axial vector mesons are expected to have close
masses. Therefore the observed states can in principle be mixtures of
$q\bar q$ and tetraquark states.  There are also possible experimental
candidates for the axial vector $1^{+-}$ $\{qs\}\{\bar q\bar
  s\}$ tetraquark with isospin $I=1$, $b_1(1960)$, and with $I=0$, $h_1(1965)$, as
well as for the $1^{+-}$ $\{ss\}\{\bar s\bar s\}$ tetraquark, $h_1(2215)$.

The  diquark-antidiquark ground state tetraquark, 
composed from an axial vector diquark and antidiquark ($A\bar A$), 
can also
be in the tensor $2^+$ state. The possible experimental candidates are
the following: a $\{qq\}\{\bar q\bar q\}$ tetraquark with $I=0$, $f_2(1910)$ or
$f_2(1950)$; the $\{qs\}\{\bar q\bar s\}$ tetraquark with $I=1$,
$a_2(1990)$ or $a_2(2080)$, and with $I=0$, $f_2(2010)$ or $f_2(2140)$;
for the $\{ss\}\{\bar s\bar s\}$ tetraquark $f_2(2340)$; for the $\{qq\}\{\bar
q\bar s\}$ tetraquark $K_2^*(1980)$.
 
There remains the important problem of describing simultaneously the
mass spectrum and decay rates of the light and heavy nonets of scalar
mesons within the relativistic quark model. This requires the
inclusion of instanton-induced mixing terms \cite{maiani} and will be
investigated in future.

In summary, we calculated the masses of the ground state light
tetraquarks in the diquark-antidiquark picture.  In distinction with
previous phenomenological treatments, we used the dynamical approach
based on the relativistic quark model. Both diquark and tetraquark
masses were obtained by numerical solution of the quasipotential wave
equations. The diquark structure was taken into account by 
using diquark-gluon form factors in terms of
diquark wave functions. It is important to emphasize  
that, in our analysis, we did not introduce any free adjustable
parameters but used their values fixed from our previous considerations
of  hadron properties. It was found that the lightest
scalar mesons $f_0(600)$ ($\sigma$), $K^*_0(800)$ ($\kappa$), 
$f_0(980)$ and $a_0(980)$ can be naturally described in our model as 
diquark-antidiquark bound systems. 

The authors are grateful to S. Gerasimov, H. Forkel, V. Matveev,
V. Savrin and D. Shirkov for support and discussions.  One of us
(V.O.G.) thanks M. M\"uller-Preussker and the colleagues 
from the particle theory group for kind hospitality. 
This work was supported in part by 
Deutscher Akademischer Austauschdienst (DAAD) (V.O.G.),
the Russian Science Support Foundation 
(V.O.G.) and the Russian Foundation for Basic Research (RFBR), grant
No.08-02-00582 (R.N.F. and V.O.G.).

\end{document}